\documentclass{ws-procs975x65}
\usepackage{graphicx}
\usepackage{bm}  
\usepackage{feynmf}
\usepackage{epsf}
\usepackage{amsmath}
\usepackage{amsfonts}
\newcommand{\beq}{\begin{equation}}
\newcommand{\eeq}{\end{equation}}
\newcommand{\bea}{\begin{eqnarray}}
\newcommand{\eea}{\end{eqnarray}}

\begin{document}

\title{An Effective Theory for the
Four-Body System with Short-Range Interactions}

\author{L.~Platter\footnote{current address \uppercase{D}epartment of
\uppercase{P}hysics and \uppercase{A}stronomy, \uppercase{O}hio
\uppercase{U}niversity, \uppercase{A}thens, \uppercase{OH} 45701,
\uppercase{USA}}}

\address{Helmholtz-Institut f\"ur Strahlen- und Kernphysik (Theorie),\\
Universit\"at Bonn, Nu\ss allee 13-15, 53113 Bonn, Germany}

\maketitle

\abstracts{We discuss the application of an effective
theory with contact interactions to the non-relativistic
four-body system with short-range interactions. We have computed
the binding energies of the $^4$He tetramer and the $\alpha$-particle.
A well-known linear correlation between three- and four-body
binding energies can be understood as a result of the absence
of a four-body force at leading order within this framework.
Further, we give results for the binding energies of the
four-boson system in two spatial dimensions.}

\section{Introduction}
Effective theories provide a convenient tool to analyze the low-energy
properties of systems characterized by a separation of scales.
If the scattering length $a$ of two non-relativistic particles
is much larger than the typical low-energy scale $l$ of the
system, one can use an effective theory with contact interactions
only to compute low-energy observables in an expansion in $l/a$.
A particular feature of this theory is that a three-body force
is needed at leading order for consistent renormalization once 
one considers systems with more than two particles. Its
renormalization group flow is governed by a limit cycle 
\cite{Bedaque:1998kg,Bedaque:1998km}.
Once this three-body is force introduced this EFT gains predictive
power and can be used to calculate three-body observables.
In \cite{Platter:2004qn,Platter:2004zs,Platter:2004ns}
we have applied this effective theory to the four-body system with
short-range interactions. Here we will summarize the most important results
without going into the technical details of the calculations.
\section{The Two- and Three-Body System}
Using an effective potential approach we are able to use
familiar quantum mechanical equations to solve for observables. 
The bare EFT potential in the two-body $S$-wave sector at leading order
is given by
\beq
\langle {\bf k}|V_{EFT}|{\bf k'}\rangle=\lambda_2~.
\eeq
This potential has to be regulated to handle appearing divergences
and to separate the low-momentum degrees of freedom from
the high-momentum degrees of freedom. The regulated potential
at leading order is $\langle {\bf k}|V|{\bf k'}\rangle = g({\bf k})\,\lambda_2\, g({\bf k'})
$,
where $g({\bf k})=\exp(-k^2/\Lambda^2)$ is the regulator function suppressing
momenta with $k\ll\Lambda$.
The resulting interaction is separable which allows to solve
the two-body problem exactly. The two-body t-matrix can be
written as
\begin{equation}
t(z)=|g\rangle\tau(z)\langle g|~,
\end{equation}
here $\tau$ denotes the two-body propagator
\begin{equation}
\tau(z)^{-1}=\frac{1}{\lambda_2}-4\pi\int\hbox{d}q \:\frac{q^2 \exp(-2q^2/\Lambda^2)}{z-q^2}
\end{equation}
Here and in the following we use units with $\hbar=m=1$. The
coupling constant $\lambda_2$ can be fixed by matching either
to the binding energy or the scattering length of the two-body
system. The three-body force (which is needed to renormalize the three-body
system) is given by
$
V_{3}=|\xi\rangle\lambda_3\langle\xi|~,
$
where $\xi({\bf u_1,u_2})=
\exp(-(u_1^2+{\textstyle \frac{3}{4}}u_2^2)/\Lambda^2)$ is the
corresponding regulator function and ${\bf u_2}$ represents the
second Jacobi momentum in the three-body system. The three-body
coupling constant $\lambda_3$ can be fixed by demanding that the
binding energy of the shallowest three-body bound state stays
constant as the cutoff $\Lambda$ is changed.\\
Various three-body observables have been calculated
using an equivalent field-theoretical approach
\cite{Bedaque:1998kg,Bedaque:1998km,Braaten:2004rn}, the
charge form factor and radius has been computed using
an effective potential approach \cite{Platter:2005sj}.
\section{The Four-Body System}
\begin{figure}[b]
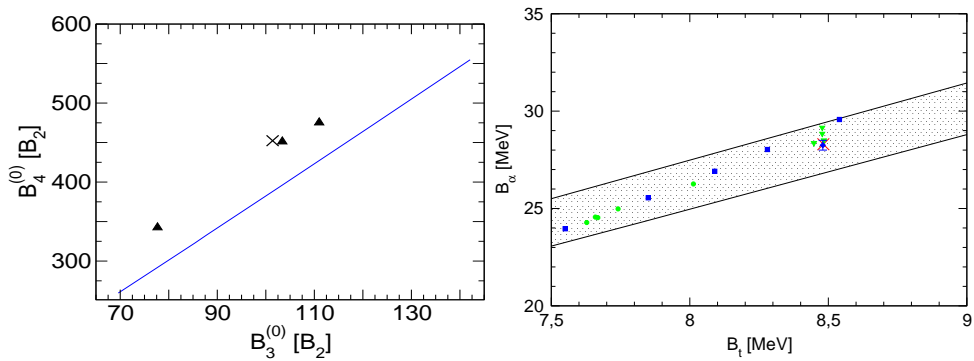

\centerline{\includegraphics*[width=2.5in,height=1.8in,angle=0]{tjon_40_2.eps}
\includegraphics*[width=2.5in,angle=0]{tjonalpha2.eps}}
\caption{\label{fig:tjonline}
The left plot shows the correlation between
the ground state energies of the $^4$He trimer and tetramer. 
The solid line shows the leading order 
effective theory result and the cross denotes the calculation for
the LM2M2 potential by Blume and Greene {\protect\cite{Blume:2000}}. The triangles
show the results for the TTY, HFD-B, and HFDHE2 potentials
{\protect \cite{Lew97,Naka83}}.
The right plot shows the correlation between the triton and the
$\alpha$-particle binding energies.
The lower (upper) line
shows our leading order result using  $a_s$ and $B_d$
($a_s$ and $a_t$) as two-body input. The grey dots and triangles show
various calculations using phenomenological potentials
without or including three-nucleon forces, respectively
{\protect \cite{Nogga:2000uu}}.
The squares show the results of chiral EFT at NLO for different cutoffs
while the diamond shows the N$^2$LO
result{\protect\cite{Epelbaum:2002vt,Epelbaum:2000mx}}.
The cross shows the experimental point.}
\end{figure}

We have computed the binding energies of the $^4$He tetramer
and the $\alpha$-particle by employing the Yakubovsky
equations. As three-body input we used the
energy of the shallowest three-body bound state as calculated
by Blume and Greene (BG) \cite{Blume:2000} and the triton binding
energy, respectively. An analysis
of the cutoff dependence of the results for the binding energies
shows that no four-body force is needed to renormalize
the four-body sector.
Further, a comparison of our $^4$He tetramer results with the values obtained by BG
\cite{Blume:2000} is shown in Table \ref{tab:results}.
The results of their calculation for the trimer and tetramer are given in the 
two right columns of Table~\ref{tab:results}, while our results are given in 
the two left columns.
In general, our results are in good agreement with the values of BG.
We also analyzed the correlation between three-body and four-body
binding energies.\\
\begin{table}[t]
\tbl{
\label{tab:results}Binding energies of the $^4$He
trimer and tetramer in mK.
The two right columns show the results by Blume and Greene
{\protect\cite{Blume:2000}} (denoted by the index BG)
while the two left columns show our results.
The number in brackets was
used as input to fix $\lambda_3$.}
{\begin{tabular}{|c||c|c||c|c|}
\hline
system & $B^{(0)}$ [mK] & $B^{(1)}$ [mK] & $B_{\rm BG}^{(0)}$ [mK] 
& $B_{\rm BG}^{(1)}$ [mK] \\ \hline\hline
$^4$He$_3$ & 127 & [2.186]   & 125.5  & 2.186\\
$^4$He$_4$ & 492 & 128  & 559.7  & 132.7\\ \hline
\end{tabular}}
\end{table}
For the $\alpha$-particle binding energy, we find 
$B_\alpha = 29.5$~MeV. If the deuteron binding energy $B_d$
 is used as input instead of the triplet scattering
length $a_t$, we obtain $B_\alpha = 26.9$ MeV. This variation 
is consistent with the expected 30\% accuracy of a leading 
order calculation in $\ell/|a|$. Our results agree
with the (Coulomb corrected) experimental value 
$B_\alpha^{exp}=(29.0 \pm 0.1)$~MeV to within 10\%.\\
In addition, our framework allows for an analysis of
universal properties of these systems by keeping
observables in the two-body system fixed and changing
the properties of the three-body system. The result of
this is shown in Fig. \ref{fig:tjonline}. The binding energies of the
three-body system are linearly correlated with the
binding energies of the four-body system. This phenomenon
has been known for a long time in nuclear physics (Tjon-line).
It follows immediately from the absence of a four-nucleon
force at leading order in this approach.

\begin{figure}[tb]
\centerline{\includegraphics*[width=3.in,angle=0]{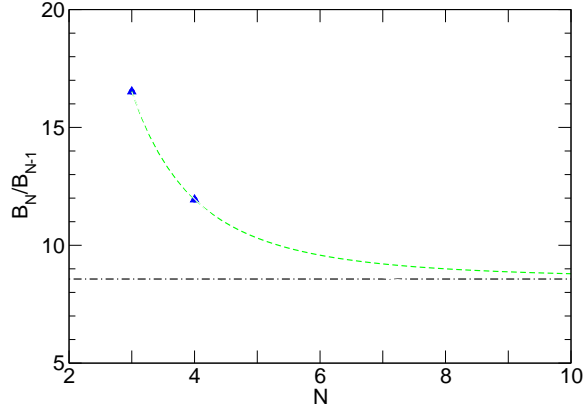}}
\caption{\label{fig:bind}
$B_N/B_{N-1}$ as a function of $N$. The dot-dashed line shows the
asymptotic value of $8.567$. The dashed line is an estimate of how
the large-$N$ value is approached.
}
\end{figure}

Low-dimensional systems are center of interest in various fields
of physics. By relying on the asymptotic freedom of
non-relativistic bosons in 2D with 
attractive short-range interactions, some 
interesting universal properties of self-bound $N$-boson droplets
were predicted for $N\gg 1$ \cite{Hammer:2004as}.
In particular, their ground state energies $B_N$ increase exponentially with $N$:
\beq
\label{BNratio}
\frac{B_{N}}{B_{N-1}}\approx 8.567 \,, \quad N\gg 1 \, .
\eeq
In light of this renewed interest, it is worthwile to calculate
the binding energies of these low-dimensional systems.
The binding energies of the three-body system were first calculated
Bruch and Tjon in 1979 \cite{Bruch}. More recently, they were
recalculated with higher precision \cite{Nielsen,Hammer:2004as}: 
the ground state has
$B_3^{(0)}=16.522688(1)\, B_2$ and there is one excited state with
$B_3^{(1)}=1.2704091(1)\, B_2$. The value of the ground state energy
$B_3^{(0)}$ differs from the prediction in Eq.~(\ref{BNratio}) by a factor
of two, indicating that the three-body system is not in the asymptotic regime.
We have calculated the binding energies of the four-boson system
in 2D using the same approach as in three spatial dimensions.
In contradistinction to the 3D case no three-body force and therefore
no three-body input is needed.
Similar to the three-body system, we find
exactly two bound states: the ground state with $B_4^{(0)}=197.3(1)\,B_2$ 
and one excited state with $B_4^{(1)}=25.5(1)\,B_2$.
We now compare our results with the large-$N$ prediction $B_N/B_{N-1}
\approx 8.567$ from Ref.~\cite{Hammer:2004as}. Using the three-body results
mentioned above, we find $B_4^{(0)}/B_3^{(0)}=11.94$.
This number is considerably closer to the asymptotic value for $B_N/B_{N-1}$ 
than the value for $N=3$: $B_3^{(0)}/B_2=16.52$. 
The exact few-body results for $N=3,4$ are compared
to the asymptotic prediction for $B_N/B_{N-1}$  
indicated by the dot-dashed line in Fig.~\ref{fig:bind}.
The dashed line gives an estimate of how the 
large-$N$ value should be approached. This estimate assumes
the expansion:
\beq
B_N = \left(c_0 +\frac{c_{-1}}{N}+\frac{c_{-2}}{N^2}
+\ldots\right)\, 8.567^N \,,
\label{eq:BNexp}
\eeq
leading to 
\beq
\frac{B_N}{B_{N-1}}= 8.567 +{\mathcal O}\left(N^{-2}\right)\,.
\label{eq:BNratioexp}
\eeq
The dashed line in  Fig.~\ref{fig:bind} was obtained by fitting the
coefficients of the $1/N$ and $1/N^2$ terms in Eq.~(\ref{eq:BNratioexp})
to the data points for $N=3,4$.
A scenario where the corrections to Eq.~(\ref{eq:BNratioexp}) already start
at ${\mathcal O}(N^{-1})$ is disfavored by the data points in Fig.~\ref{fig:bind}.

\section{Summary}
We have shown, that four-body systems with large scattering length
can be described within the framework of an effective theory
with contact interactions only. Our results for the binding
energies of the $^4$He tetramer and the $\alpha$-particle
are in good agreement with theoretical calculations and the
experimental value, respectively. The linear correlation between
the three-body binding energies and the four-body binding energies
(Tjon-line)
turns out to be a universal result of systems with large scattering
length and can be understood by the absence of a four-body force
at leading order.\\
Further, we computed the binding energies of the four-boson system
in two spatial dimensions and showed that the four-body results
lie significantly closer to the assymptotic value for $B_N/B_{N-1}$
than the three-body results.
\section*{Acknowledgments}
This work was done in collaboration with H.-W.~Hammer and U.-G.~Mei\ss ner. 
This research was supported in part by the the Deutsche Forschungsgemeinschaft
through funds provided to the SFB/TR~16 and by
the EU Integrated Infrastructure Initiative Hadron Physics.

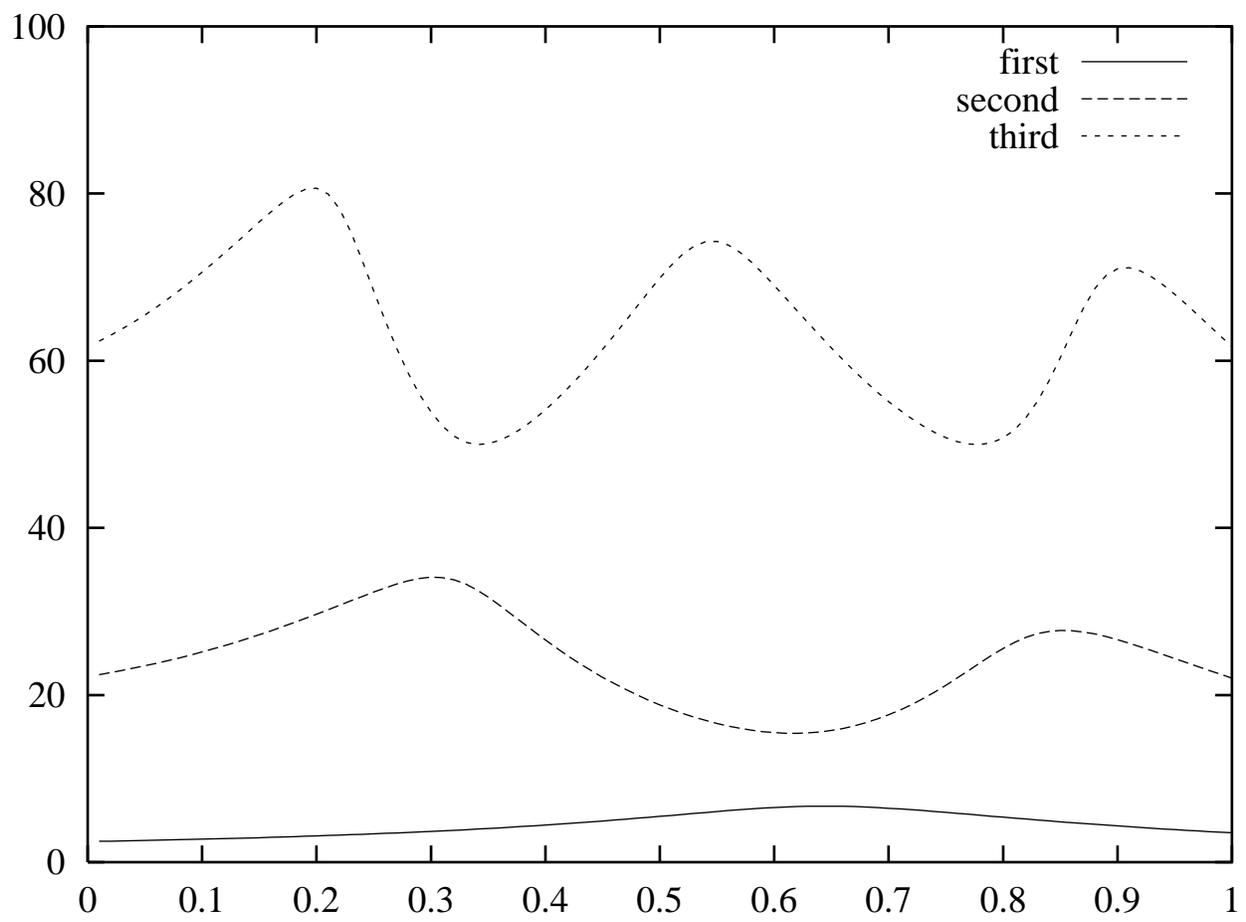
\end{document}